\documentclass[aps,prm,groupedaddress,reprint,showpacs,floatfix,pdftex]{revtex4-2}
\usepackage{graphicx}
\usepackage{comment}
\usepackage{tabularx}
\usepackage{dcolumn}
\usepackage{bm}
\usepackage{amsmath}
\usepackage{amssymb}
\usepackage{txfonts}
\usepackage{url}
\usepackage[dvipdfmx]{color}
\usepackage{multirow}
\usepackage{ulem}
 
\newcommand{\tadd}[1]{{\color{black}#1}}

\newcommand{\ghost}[1]{}

\begin{document}
\title{A New Candidate Structure of the H$_2$-PRE Phase of Solid Hydrogen}
\author{Tom Ichibha$^\dagger$}
\email[]{ichibha@icloud.com}
\affiliation{Materials Science and Technology Division, Oak Ridge National Laboratory, Oak Ridge, Tennessee 37831, USA}
\author{Yunwei Zhang}
\email{These two authors made the same contribution.}
\affiliation{School of Physics, Sun Yat-sen University, Guangzhou 510275, China}
\affiliation{Cavendish Laboratory, University of Cambridge, Cambridge CB30HE, United Kingdom}
\author{Kenta Hongo}
\affiliation{Research Center for Advanced Computing Infrastructure, JAIST, Asahidai 1-1, Nomi, Ishikawa 923-1292, Japan}
\author{Ryo Maezono}
\email[]{\tadd{rmaezono@me.com}}
\affiliation{School of Information Science, JAIST, Nomi, Ishikawa 923-1292, Japan}
\author{Fernando A. Reboredo}
\email[]{\tadd{reboredofa@ornl.gov}}
\affiliation{Materials Science and Technology Division, Oak Ridge National Laboratory, Oak Ridge, Tennessee 37831, USA }
%
\date{\today}
\begin{abstract}
  \tadd{
    Experimental progress finally reached the metallic solid hydrogen phase,
    which was predicted by Wigner and Huntington over 80 years ago.
    However, the different structures in the phase diagram are still been debated
    due to the difficulty of diffraction experiments for high-pressured hydrogen.
    The determination of crystal structures under extreme condition is both of  the basic condensed matter physics,
   and in planetary science:  
    the behavior of giant gaseous planets (e.g. Jupiter, Saturn...) strongly depends on the properties of
    inner high-pressured hydrogen.
  }
  This work describes new possible structures appearing under high pressures of 400$\sim$600 GPa. 
  We applied a structural search using particle swarm optimization with density functional theory (DFT) 
  to propose several candidate structures. For these structures, we performed fixed-node diffusion Monte Carlo simulations 
  combined with DFT zero-point energy corrections to confirm their relative stability.
  \tadd{
    We found $P2_{1}/c$-8 as a promising candidate structure for the H$_2$-PRE phase.
    $P2_{1}/c$-8 is predicted the most stable at 400 and 500~GPa. $P2_{1}/c$-8 reproduces qualitatively the IR spectrum peaks observed in the H$_2$-PRE phase.
  }
  \footnote{
  This manuscript has been authored by UT-Battelle, LLC, under contract DE-AC05-00OR22725
  with the US Department of Energy (DOE). The US government retains and the publisher,
  by accepting the article for publication, acknowledges that the US government retains a nonexclusive,
  paid-up, irrevocable, worldwide license to publish or reproduce the published form of this manuscript,
  or allow others to do so, for US government purposes. DOE will provide public access to these results
  of federally sponsored research in accordance with the DOE Public Access Plan
  (http://energy.gov/downloads/doe-public-access-plan).
  }
\end{abstract}
\maketitle

\section{Introduction}
\label{sec.intro}
Dense hydrogen undergoes phase transitions to solid crystals under high pressure, 
providing a variety of solid structural phases. 
At the lowest pressure, hydrogen assumes a hexagonal lattice of molecules 
with sufficient internal degrees of freedom to rotate each molecule (phase I)~\cite{1996LOU}. 
At around 110~GPa, the rotation freezes to a uniform orientation (phase II)~\cite{1996LOU}. 
\tadd{
  At higher pressures, Raman and infrared (IR) spectrum experiments have revealed 
  the existence of six distinct structural phases (Phases III~\cite{1994MAO}, IV, IV'~\cite{2012HOWa,2012HOWb,2016DAL}, 
  V~\cite{2016DAL}, H$_2$-PRE~\cite{2019RPD_IFS}, and metallic hydrogen (MH)~\cite{2016DIA,2017RPD_IFS,2020PL_PD}).
  (Table I of \cite{2019RPD_IFS} provides a complete summary of the structural phases and transition pressures.
  However, another experimental work \cite{2020PL_PD} proposed another phase diagram, which does not include the H$_2$-PRE phase.)    
}
However, it is technically difficult to identify the structural characteristics 
of these phases using X-ray or neutron diffraction experiments~\cite{1996LOU}. 
Hence, elucidation of these structures has been left as an important mission 
for electronic structure calculations. 

\vspace{2mm}
Pioneering work~\cite{2000JOH}, molecular dynamics simulation at a fixed pressure~\cite{1994GLE} 
at the phase-III range ($>$150GPa), was performed to predict the $Cmca$-4 structure. 
Subsequently, an {\it ab initio} random-search simulation~(AIRSS) ~\cite{2007PIC} was applied 
to find several additional candidate structures for phase III. 
Although this included the $Cmca$-4 structure, it was claimed that $C2/c$-24 was most likely, 
because its Raman spectrum was found to well-reproduce the experimentally observed frequencies~\cite{2007PIC}.
Subsequent AIRSS investigations have reported more energetically stable structure than $C2/c$-24:
a hexagonal $P6_1$22 appearing at a phase-III pressure range ($>$150GPa)~\cite{2016MON}.
The Raman spectrum of the $P6_1$22 structure is consistent with experimental observations ~\cite{2016MON}.
Further experimental investigations on the pressure dependence of the spectrum ~\cite{2017AKA}
has revealed a structural transition at around 190~GPa. 
Powder X-ray diffraction experiments have confirmed that this transition is accompanied
by the deformation of an HPC-like structure, which is consistent with theoretical predictions 
regarding the transition from $P6_1$22 $\to$ $C2/c$-24 (deformed from six-fold symmetry)~\cite{2016MON}.
\tadd{
  A structure originally found as a candidate structure of the phase III, 
  $Cmca$-12~\cite{2007PIC}, was also recently focused on as a candidate structure of the H$_2$-PRE phase~\cite{2019RPD_IFS}.
  This phase was observed between the Phases III and MH~\cite{2019RPD_IFS}. 
  The transition pressures are 360 and 495~GPa~\cite{2019RPD_IFS}.
  The $Cmca$-12 structure was predicted the most stable from 424(3) to 447(3) GPa by quantum Monte Carlo,
  between the candidate structures of phases III and MH~\cite{2015MCM}. 
  In addition, the IR spectrum of $Cmca$-12 qualitatively agrees with the experiment~\cite{2019RPD_IFS,2007PIC}.
}

\vspace{2mm}
\tadd{
  While the phases I--III, H$_2$-PRE, and MH are stable down to near 0~K, 
  the phases IV and IV' are stable around or above room temperature~\cite{2012HOWa,2012HOWb}.
}
The transition from  phases III to IV occurs around 220~GPa~\cite{2012HOWa}.
From the Raman spectrum, the structure of phase IV was expected to be a mixed structure of atomic and molecular crystals. 
It was also found that the band gap and the Raman spectrum showed discontinuous behaviors at 275~GPa, 
implying a phase transition from IV to IV'~\cite{2012HOWb}.
AIRSS was used to predict phases IV and IV'. Phase IV was proposed to be a mixed structure ($Pc$-48),
where two inequivalent layers alternatively appear.
One layer is formed by strongly-bound molecules, and the other layer is formed by a graphene-like array of atomic hydrogen\cite{2012PIC}. 
This mixed structure was found to be energetically more stable than others proposed thus far at room temperature
in the range of $250\sim 295$~GPa~\cite{2012PIC}. 
The computed Raman spectrum of the structure was also found to be consistent with experiments~\cite{2012PIC}.

\begin{table}[htbp]
  \begin{center}
    \caption{
      \label{tab.summary} \ghost{tab.summary}
      Most promising candidates of each structural phase.
    }
    \begin{tabular}{cccc}
      \hline
      Phase & Structure & Description & Reference \\
      \hline
      II       & $P2_{1}/c$ & HCP\footnote[1]{hexagonal close-packed} of molecules & \cite{2007PIC} \\
      &          $P6_{3}/m$ & HCP\footnotemark[1] of molecules & \cite{2007PIC} \\
      III (LP)\footnote[2]{low pressure side of phase III} & $P6_{1}22$ & layered molecular & \cite{2016MON} \\
      III (HP)\footnote[3]{high pressure side of phase III} & $C2/c$-24  & layered molecular & \cite{2007PIC} \\
      IV/IV'   & $Pc$-48    & mixed crystal & \cite{2012PIC} \\
      V        & $Pca2_{1}$ & mixed crystal & \cite{2018MON} \\
      \tadd{H$_2$-PRE} & \tadd{$Cmca$-12} & \tadd{molecular crystal} & \tadd{\cite{2007PIC,2015MCM}} \\
      \hline
    \end{tabular}
  \end{center}
\end{table}

\vspace{2mm}
By further increasing the pressure at room temperature, another phase V structure was discovered under $\sim$325~GPa~\cite{2016DAL}. 
Its Raman spectrum suggests that the band gap shrinks while the molecules are dissociated as the pressure increases.
\tadd{Therefore, this phase was considered as a stepping stone to the MH phase.}
The spectrum cannot be explained by any \tadd{of the} proposed structures \tadd{at that time}, 
stimulating further structural searches beyond the harmonic approximation~\cite{2018MON}. 
The anharmonic approach~\cite{2018MON} was then used to discover a new $Pca$2$_1$ structure, 
which provided a spectrum that was consistent with the experiments. 
The structure consists of hydrogen molecules \tadd{with} longer bond lengths 
that lead to narrower band gaps, which are consistent with the features experimentally captured in phase V. 
Table \ref{tab.summary} summarizes the most promising candidates, in our view, of each structural phase predicted theoretically. 

\vspace{2mm} 
In this study, we perform a structural search at zero temperature 
in a high-pressure region ($400 \sim 600$~GPa), 
where a structure has yet to be \tadd{determined by experiments}. 
By using the particle-swarm optimization (PSO) algorithm implemented 
in CALYPSO~\cite{2010WAN,2012WAN}, we found 10 new candidate structures 
beyond those reported in existent publications 
($C2/c$-24\cite{2007PIC}, $Cmca$-12\cite{2007PIC}, $Cmca$-4\cite{2000JOH,2007PIC}, 
$I4_1/amd$\cite{2011MCM}, and $mC$-24\cite{2012LIU}). 
For these candidates, we performed fixed-node diffusion Monte Carlo (FNDMC) calculations~\cite{2001FOU}.
FNDMC includes many body corrections beyond DFT, to obtain static formation enthalpies. 
This reliable framework has been already applied in several previous 
solid hydrogen publications~\cite{2014AZA,2015MCM,2015DRU,2018MON,2020AZA,1987CEP}. 
\tadd{
  Note that, in addition, other QMC methods (i.e., path integral and reputation QMC) have been also utilized
  to study the properties of fluid hydrogen \cite{2020GOR, 2019RIL, 2018PIE, 2013MORa, 2017PIE, 2016PIE}.
}

\vspace{2mm}
\tadd{
  Among the structures found by structural search, $Pbam$-8 is predicted by FNDMC to have a 
  static enthalpy comparable with the structures reported 
  in the previous studies as shown in Table \ref{tab.summary}. 
  However $Pbam$-8 has imaginary frequency phonons which signals a structural instability.  
  Therefore, we relax the structure along the direction of 
  the imaginary mode and obtain a slightly different structure, $P2_{1}/c$-8. 
  We find that $P2_{1}/c$-8 has the lowest dynamic enthalpy at zero temperature and at 400~GPa and 500~GPa. 
  The dynamic enthalpy is given as the sum of the zero-point energy (ZPE) predicted by DFT and FNDMC static enthalpy. 
  Therefore, we propose $P2_{1}/c$-8 to be a new candidate structure of the H$_2$-PRE phase~\cite{2019RPD_IFS}. 
  We confirm that the IR spectrum of $P2_{1}/c$-8 qualitatively agrees with the experiment for the H$_2$-PRE phase, 
  as well as the conventional candidate, $Cmca$-12~\cite{2019RPD_IFS}. 
}

\section{Methods}
\label{methods}
A PSO structural search using CALYPSO~\cite{2010WAN,2012WAN} is performed 
under 400~GPa with different unit-cell sizes comprising 2 to 70 atoms. 
Forty structures are generated with every iteration, and 30~\% of the structures 
are generated using the artificial bee-colony algorithm~\cite{2016BIA} 
according to the structures appearing in the prior iteration; however, 
the remaining 70~\% are generated randomly. 
We run 45 iterations for every unit-cell size. 
We use the Perdew--Burke--Ernzerhof (PBE) DFT method implemented in the Vienna Ab initio Simulation Package (VASP)~\cite{1996KRE} 
for geometric optimization and energy evaluation. 
Ionic cores are represented by projector-augmented wave pseudo potentials~\cite{1999KRE}. 
The cutoff energy ($E_{\mathrm{cut}}$) for the plane wave basis set 
is chosen as 700~eV on the $k$-mesh for the integration over the Brillouin zone with a mesh size $<$~0.30 $\AA^{-1}$. 
The threshold for self-consistent field (SCF) convergence is taken as 1.0$\times$10$^{-5}$~eV. 
Structural relaxations are performed until the total energy converges within 
the threshold: 1.0$\times$10$^{-5}$~eV.

\vspace{2mm}
On obtaining candidate structures using CALYPSO, we replace the exchange-correlation (XC)
functionals into Van der Waals (vdW) density functionals (DFs) ~\cite{2004DIO} to perform additional
structural relaxations at 400, 500, and 600~GPa using VASP~\cite{1996KRE} with $E_{\mathrm{cut}}$ = 1,500~eV
and $k$-mesh sizes as $<$ 0.25~$\AA^{-1}$, which are the conditions needed to achieve total energy
convergence within 0.1 mHartree/atom for the $I4_1/amd$ structure.
SCF convergences are achieved using tighter thresholds of 1.0$\times$10$^{-6}$~eV, and the relaxations are
performed until the force field at each ion falls to within 1.0$\times$10$^{-3}$ eV/$\AA$.

\vspace{2mm}
ZPE is evaluated using the frozen phonon scheme implemented in Phonopy ~\cite{2015TOG} coupled
with VASP as the DFT kernel under the same computational conditions mentioned above. 
We take the simulation cell size to be larger than 72~f.u.,
for which the finite size error of ZPE is confirmed to be less than 0.1~mHartree/atom for the $I4_1/amd$ structure.
We also confirm that the ZPE hardly depends on the functional.
In addition to vdW-DF, we perform PBE \cite{1996PER} to evaluate the ZPE,
finding that their difference are just 0.03 mHartrees in the case of I4$_1$/amd.
\tadd{We used Phonopy-Spectroscopy to calculate IR spectrum~\cite{2017JMS_AW}.}

\vspace{2mm}
We evaluate FNDMC static enthalpies using QMCPACK ~\cite{2018KIM}. 
Ionic cores are represented by the effective core potential developed
for FNDMC calculations~\cite{2018ANN}.
Our fixed-node trial wave function is a Slater--Jastrow type ~\cite{2001FOU},
wherein orbital functions used to form the determinant 
are generated by DFT with vdW-DF functionals~\cite{2004DIO} 
implemented in Quantum Espresso~\cite{2009GIA}.
The cutoff energy is 300~Ry, for which the total energies of $C2/c$-24 and $I4_1/amd$
converge to within 0.04 mHartree/atom. 
The Jastrow function consists of the one-, two- and three-body terms,
amounting to 78 variational parameters. 
The parameters are optimized using the linear method~\cite{2007UMR} implemented in QMCPACK. 
We use a simulation cell comprising $\sim$100 atoms for FNDMC calculations
to suppress the two-body finite size error.
We also apply the kinetic energy correction~\cite{2006CHI} and the model periodic
Coulomb interaction~\cite{1999KEN,1996FRA,1997WIL}.
We use twist-averaged boundary conditions~\cite{2001LIN} with a 5$\times$5$\times$5 twist-grid,
with which the total energies of $C2/c$-24 
and $I4_1/amd$ are converged to within 0.1 mHartree/atom.
The timestep of FNDMC calculation is 0.02 a.u.$^{-1}$, with which the total energies
of $C2/c$-24 and $I4_1/amd$ converge to within 0.05 mHartree/atom.
For all FNDMC calculations, the target population is set to 1,024. 
The random walkers evolve during 
equilibration and statistical accumulation steps
of 2,000 and 48,000, respectively.

\vspace{2mm}
\tadd{
  In FNDMC, forces are practical for only simple small systems
  due to the computational cost. 
  The second derivatives of energy are much more difficult
  so they are not implemented yet in QMCPACK to our knowledge.  
  Without a Hessian, we cannot calculate phonons.     
  Therefore, we rely in DFT methods to obtain geometry and phonons in this work.
  We note that, to our knowledge, there is only one example of FNDMC phonon calculation for diamond~\cite{2021KN_SS}
  using TurboRVB~\cite{2020KN_SS}.   
}

\vspace{2mm}
For fairness of comparison with other studies, 
we do not perform volume optimization at the FNDMC level. An 
additional $P\cdot V$ term in the static enthalpy 
is evaluated by the $V\left(P\right)$ dependence 
using DFTs at $P=$ 400, 500, and 600~GPa. 
By adding ZPE, we obtain dynamic enthalpies.

\begin{figure}[htbp]
  \includegraphics[width=0.8\hsize]{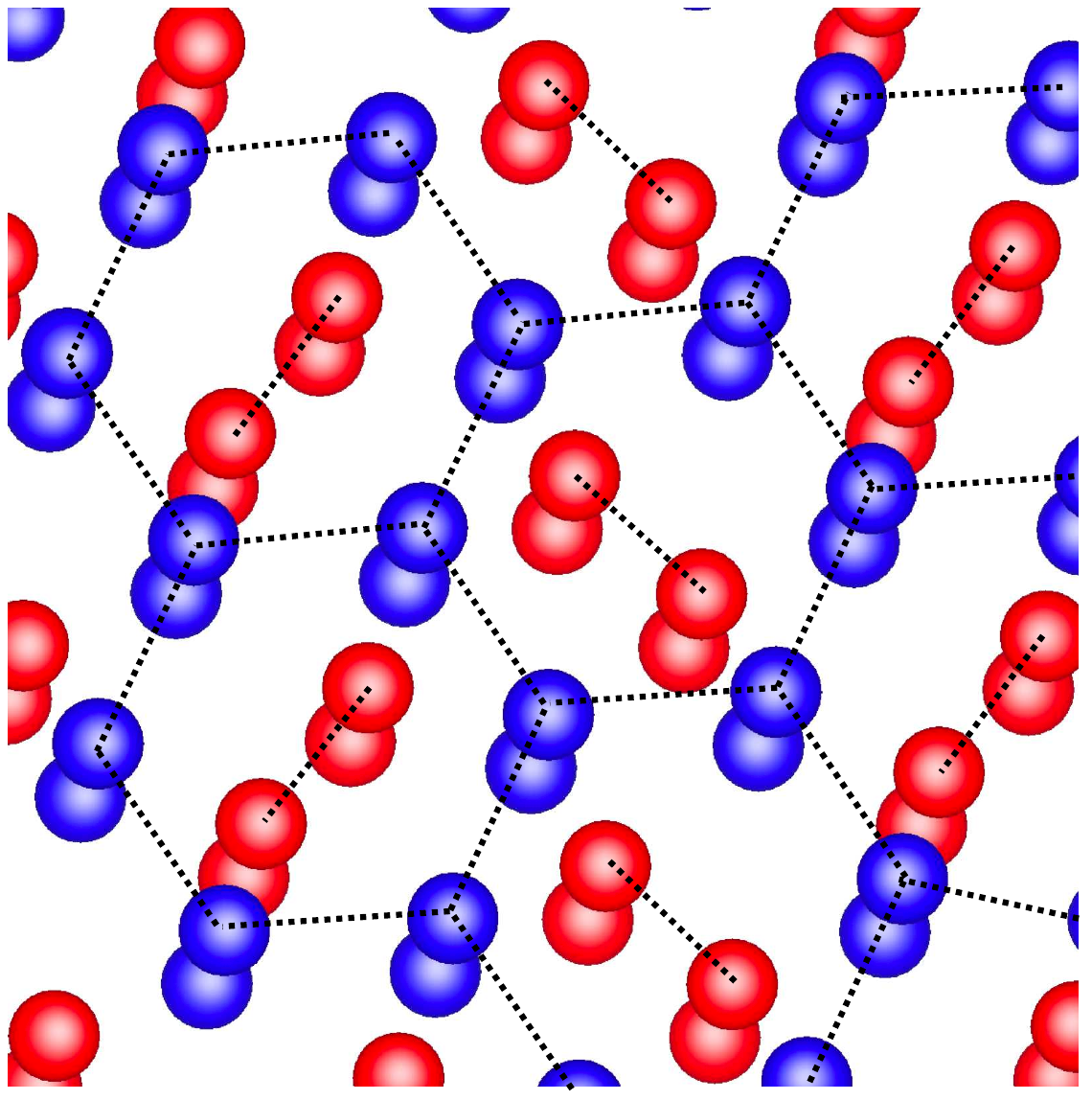}
  \caption{
    \label{fig.pbam8_str}\ghost{fig.pbam8\_str}
    The crystal structure of $Pbam$-8. 
    Hydrogen molecules with very short bond lengths 
    are located with their axes vertical to the layer. 
    The layers formed by the blue molecules and 
    red ones are alternately stacked. 
    Although the blue layer forms a honeycomb lattice, 
    the molecules in the red layer are ordered 
    such that each pair accommodated within 
    the honeycomb cell is defined by two adjacent blue layers 
    flanking the red one.
  }
\end{figure}
\begin{figure}[htbp]
  \includegraphics[width=\hsize]{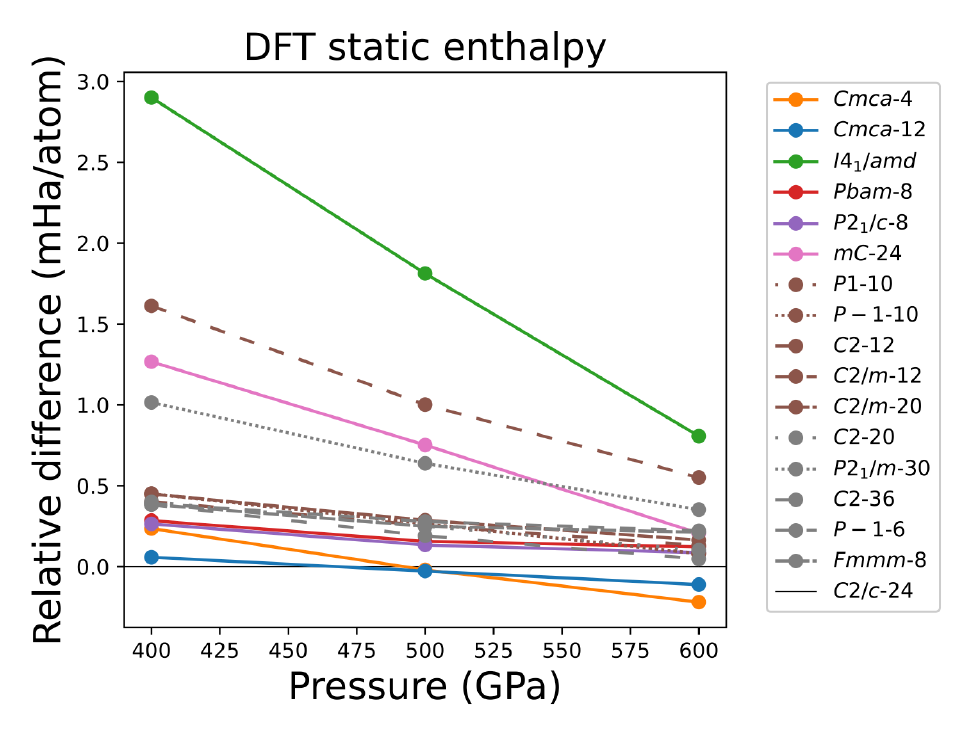}
  \caption{\label{fig.dft_static}\ghost{fig.dft\_static}
    Comparisons of static enthalpies evaluated by 
    DFT in terms of the difference from 
    the $C2/c$-24 structure (zero reference 
    in the plot). 
    The most stable structure is found to 
    be $C2/c$-24$\to Cmca$-12$\to Cmca$-4 
    as pressure increases.
    \tadd{
    Figure \ref{fig.dmc_static} shows more reliable
    prediction of static enthalpies by FNDMC.
    }
  }
\end{figure}
\begin{figure}[htbp]
  \includegraphics[width=\hsize]{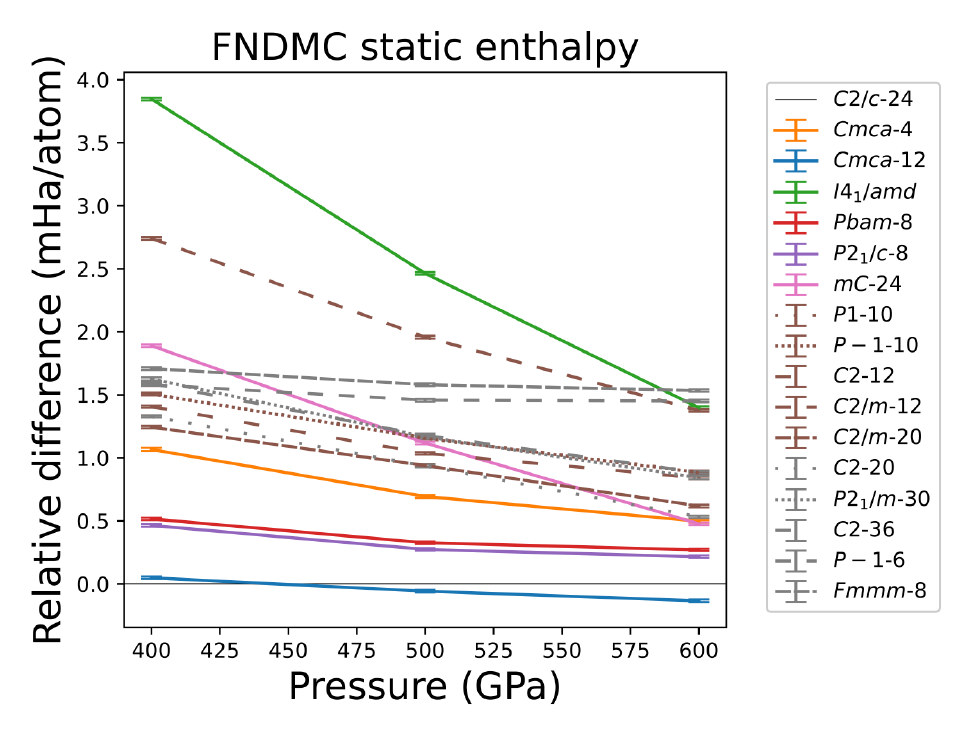}
  \caption{\label{fig.dmc_static}\ghost{fig.dmc\_static}
    Comparisons of static enthalpies evaluated by FNDMC
    in terms of the difference from the $C2/c$-24 structure
    (zero reference in the plot). 
    Statistical error bars are shown.
    \tadd{
    Figure \ref{fig.dmc_dynamic} shows dynamic enthalpies
    calculated as the sum of this FNDMC static enthalpies
    and DFT ZPE shown in Figure \ref{fig.zpe}.
    }
  }
\end{figure}

\section{Results and Discussions}
\label{sec.result}
Using the PSO structural search, we find 10 new candidate structures beyond those
reported in preceding works ($C2/c$-24, $Cmca$-12, $Cmca$-4, $I4_1/amd$, and $mC$-24). 
Comparisons of their static enthalpies predicted by vdW-DF functional
are shown in Fig.~\ref{fig.dft_static}.
The 10 structures include nine molecular crystals and one mixed structure, $Pbam$-8.
$Pbam$-8 comprises of atomic and molecular crystal layers appearing alternatively,
as shown in Figure \ref{fig.pbam8_str}.
The preceding PBE study~\cite{2012LIU, 2015MCM} reports that $mC$-24~(quasi-molecular)
and $I4_1/amd$~(atomic) structures have the lowest static enthalpy under pressures
higher than $\sim$500 GPa.
In contrast, these structures are evaluated as being relatively unstable in our vdW-DF results,
as shown in Figure \ref{fig.dft_static}.
McMinis \textit{et al.} compare static enthalpies using both vdW-DF and PBE,
reporting that atomic crystals, $I4_1/amd$ and $\beta$-Sn, are more unstable than molecular crystals,
$C2c$-24, $Cmca$-4, and $Cmca$-12, when vdW-DF is applied~\cite{2015MCM}.
Similarly, $mC$-24~(quasi-molecular) is likely to get unstable
when vdW-DF is applied, as shown in the present work, owing to the longer molecular distances.

\begin{figure}
  \centering
  \includegraphics[width=\hsize]{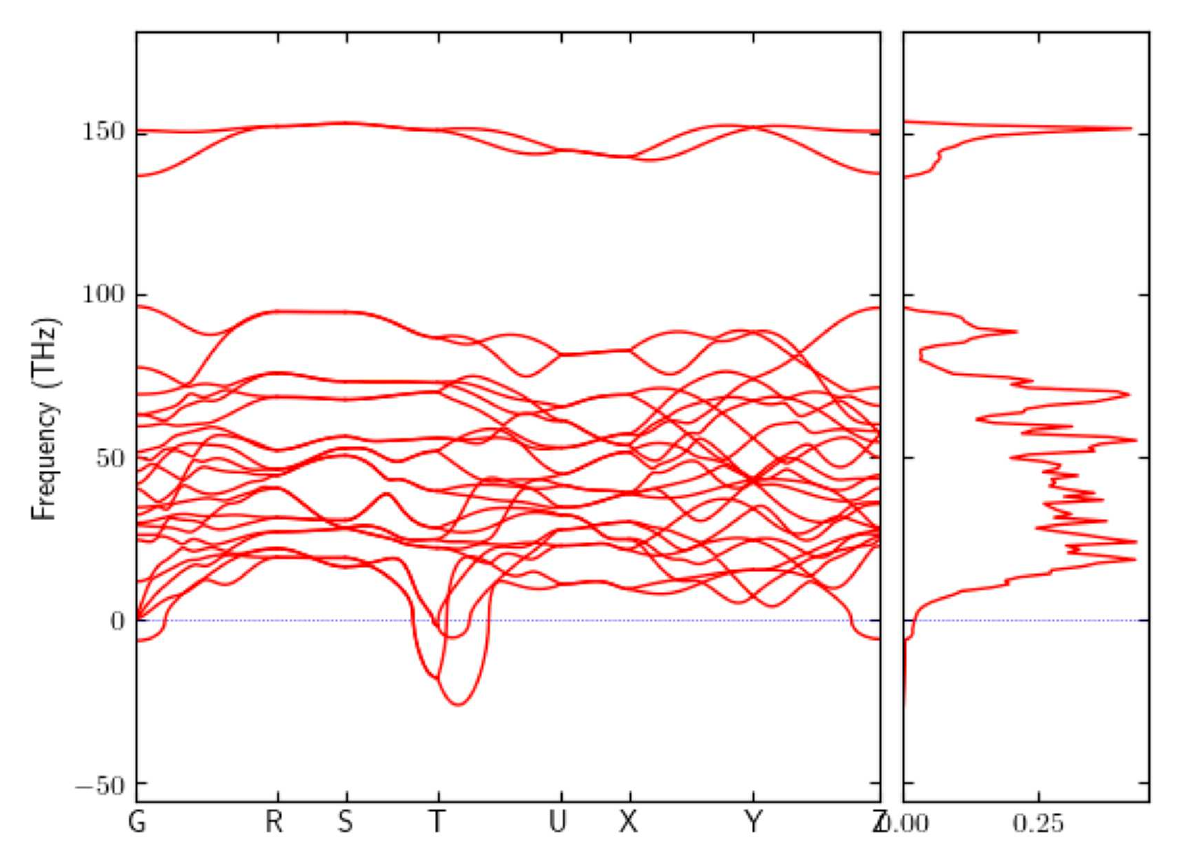}  
  \caption{
    \label{fig.pbam8_phonon}\ghost{fig.pbam8\_phonon}
    Phonon dispersion of $Pbam$-8 structure at 400~GPa.
    The correspondences between labels and $\vec q$-points are 
    G:~$\vec q$=(0.0,0.0,0.0), 
    R:~$\vec q$=(0.5,0.5,0.5), 
    S:~$\vec q$=(0.5,0.5,0.0), 
    T:~$\vec q$=(0.0,0.5,0.5), 
    U:~$\vec q$=(0.5,0.0,0.5), 
    X:~$\vec q$=(0.5,0.0,0.0), 
    Y:~$\vec q$=(0.0,0.5,0.0),
    and Z:~$\vec q$=(0.0,0.0,0.5). 
  }
\end{figure}
\begin{figure}
  \centering
  {\Large (a) side view}
  \includegraphics[width=\hsize]{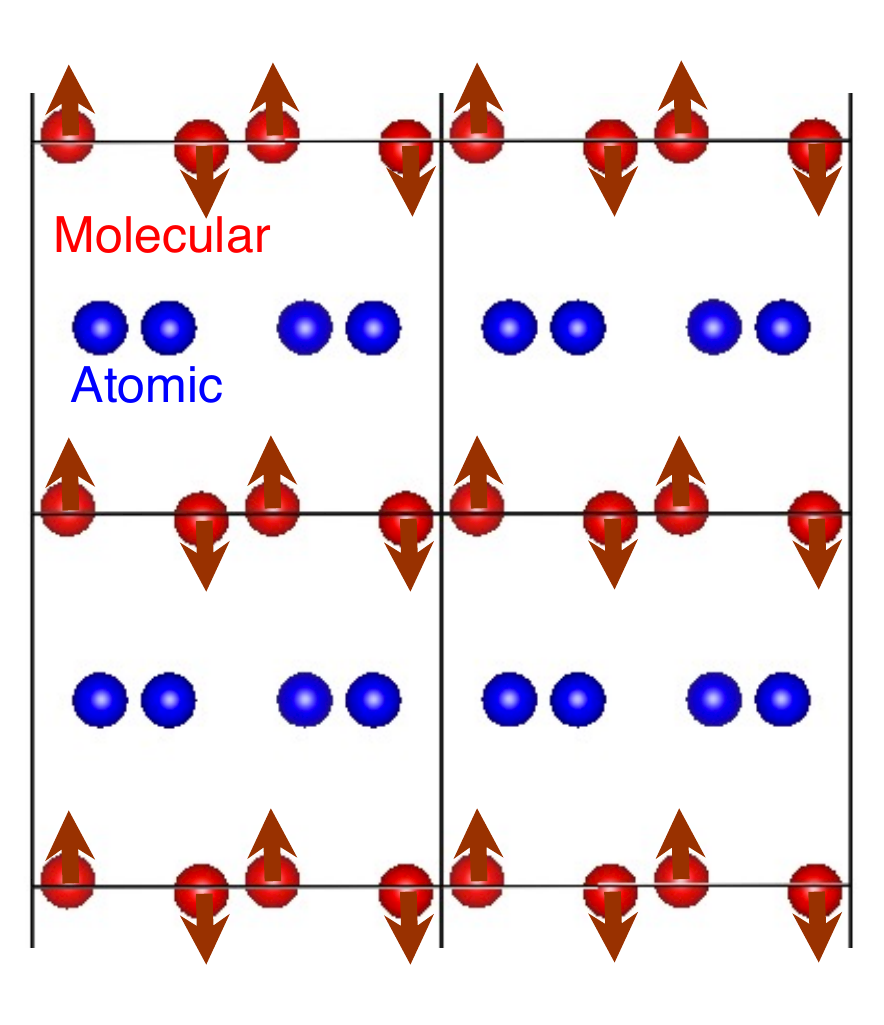}
  {\Large (b) top view}
  \includegraphics[width=\hsize]{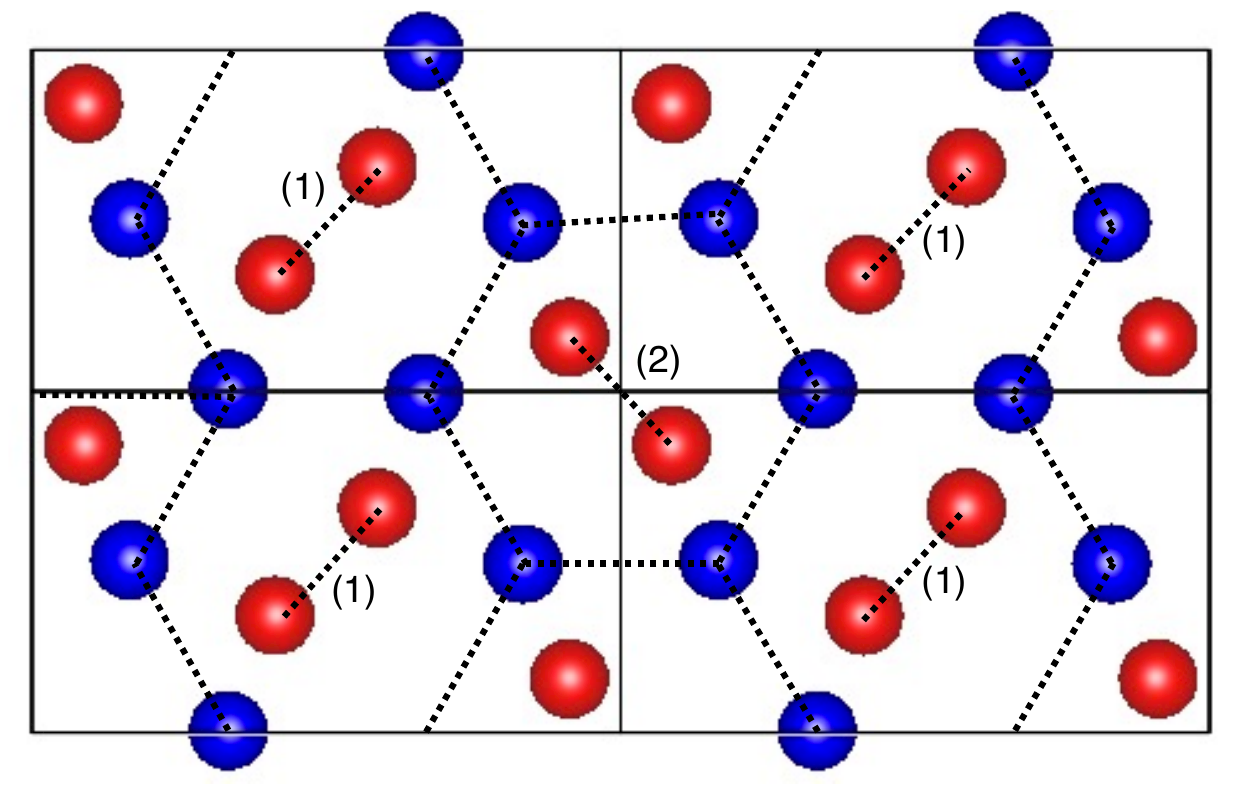}
  \caption{
    \label{fig.p21c8_structure}\ghost{fig.p21c8\_structure}
    Crystal structure of $P2_{1}/c$-8.
    The hydrogen atoms in the molecular (atomic) crystal
    layers are shown in red (blue).
    The arrows in the out-of-plane direction
    represent the shifts from $Pbam$-8.
  }
\end{figure}
\begin{figure}
  \centering
  \includegraphics[width=\hsize]{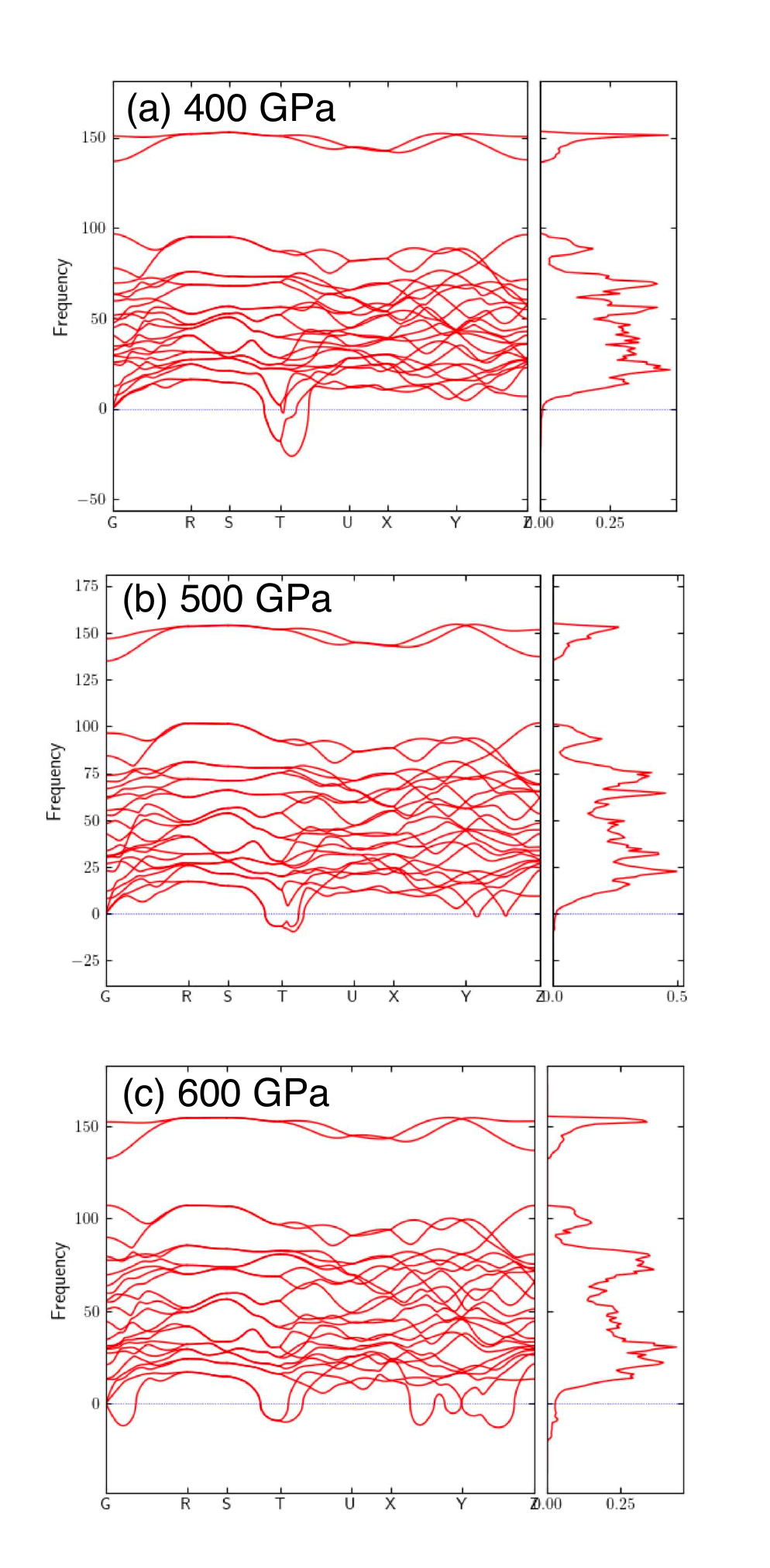}
  \caption{
    \label{fig.p21c_phonon}\ghost{fig.p21c\_phonon}
    Phonon dispersions of $P2_{1}/c$-8 structure at 400, 500, and 600~GPa.
    The correspondences between labels and $\vec q$-points are 
    $\mathrm{\Gamma}$:~$\vec q$=(0.0,0.0,0.0), 
    R:~$\vec q$=(0.5,0.5,0.5), 
    S:~$\vec q$=(0.5,0.5,0.0), 
    T:~$\vec q$=(0.0,0.5,0.5), 
    U:~$\vec q$=(0.5,0.0,0.5), 
    X:~$\vec q$=(0.5,0.0,0.0), 
    Y:~$\vec q$=(0.0,0.5,0.0),
    and Z:~$\vec q$=(0.0,0.0,0.5).     
  }
\end{figure}

\vspace{2mm}
\tadd{
  Figure \ref{fig.dmc_static} shows static enthalpies of structural phases predicted by FNDMC. 
  However, the 10 structures are predicted to have negative phonons. Therefore, we tried to find a 
  dynamically stable structure based on the phonon dispersion.
  Among the 10 structures we find, $Pbam$-8 has a notably low static enthalpy. 
  $P2_1/c$-8 was found after relaxing $Pbam$-8 in the direction of instability. 
  Figure \ref{fig.pbam8_phonon} shows the phonon dispersion and phonon density of states (phonon DOS) of $Pbam$-8 at 400~GPa.
  There are imaginary modes around the $\mathrm{\Gamma}$ point and near the T point.
  We relax the structure along the direction of the imaginary mode at the $\Gamma$ point
  and obtain a different structure, $P2_{1}/c$-8, as shown in Figure \ref{fig.p21c8_structure}.
  Compared with $Pbam-8$, the hydrogen molecules are shifted out of a plane,
  as shown by arrows in Figure \ref{fig.p21c8_structure}(a).
  The phonon dispersion and phonon DOS of $P2_{1}/c$-8 at 400~GPa are shown in Figure \ref{fig.p21c_phonon}(a).
  $P2_{1}/c$-8 does not have negative phonons around the $\mathrm{\Gamma}$ point.   
  We similarly try to relax $P2_{1}/c$-8 along the direction of the imaginary mode at T point.
  However, the relaxed structure is identical to $P2_{1}/c$-8.    
  We argue that small numerical errors in the harmonic finite difference approach implemented in Phonopy~\cite{2015TOG}
  or anharmonic effects might be responsible of those remaining imaginary frequency modes, since the structure is stable
  under deformation along those distortions. However,
  the population of imaginary modes in the harmonic phonon DOS is insignificant.
  Figure \ref{fig.p21c_phonon}(b)(c) shows the phonon dispersion and phonon DOS of $P2_{1}/c$-8 at 500 and 600~GPa.  
  The population of imaginary modes is negligible at 500~GPa but significantly large at 600~GPa, suggesting an instability there.
  Therefore, we calculated ZPE at 400 and 500~GPa only.
}

\begin{figure}[htbp]
  \includegraphics[width=\hsize]{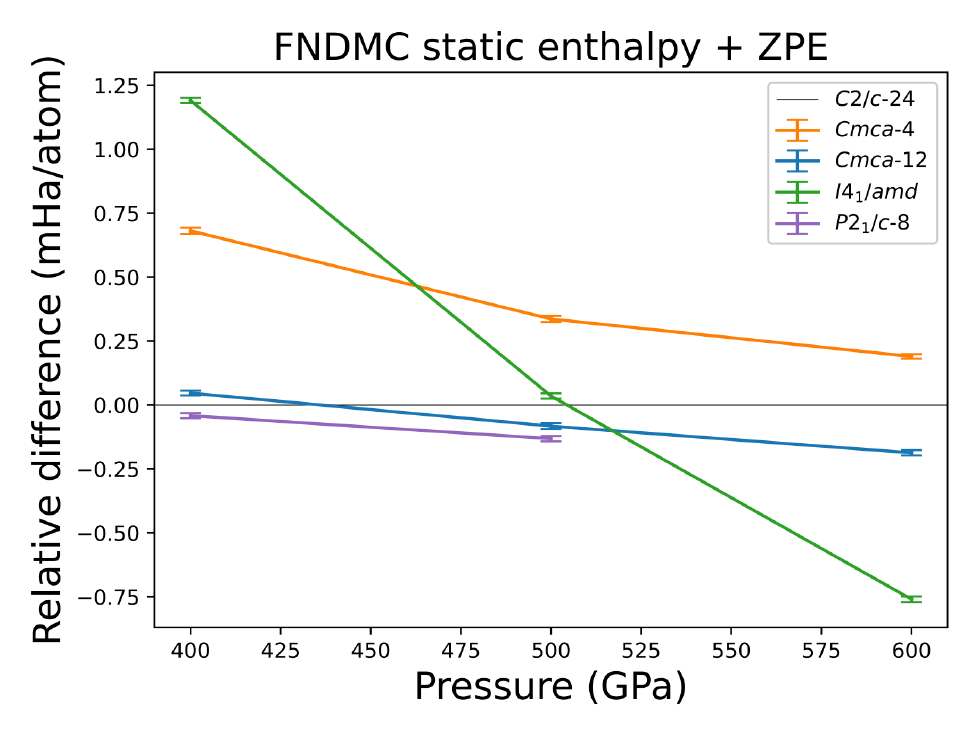}
  \caption{
    \label{fig.dmc_dynamic}\ghost{fig.dmc\_dynamic}
    Comparisons of dynamic enthalpies evaluated by 
    FNDMC and DFT phonon evaluations 
    in terms of the difference from the $C2/c$-24 structure
    (zero reference in the plot).
    \tadd{
    This is the main result of this paper.
    Statistical error bars are shown. 
    }
  }
\end{figure}
\begin{figure}[htbp]
  \includegraphics[width=\hsize]{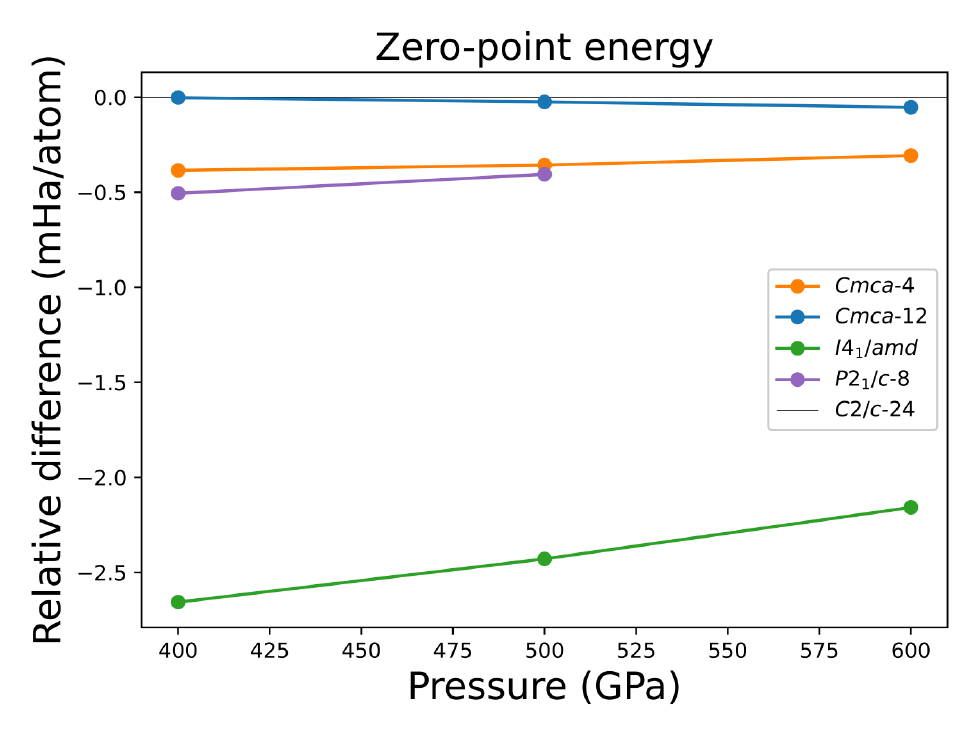}
  \caption{\label{fig.zpe}\ghost{fig.zpe}
    Comparisons of DFT ZPE predicted by the phonon calculations
    in terms of the difference from the $C2/c$-24 structure (zero reference in the plot). 
  }
\end{figure}

\vspace{2mm}
\tadd{
  Fig.~\ref{fig.dmc_dynamic} shows the dynamic enthalpies 
  (sum of the FNDMC static enthalpy; Fig.~\ref{fig.dmc_static}, 
  and the ZPE calculated by DFT-phonon calculations (Fig.~\ref{fig.zpe}).
  Among the reported structures (\textit{i.e.,} except for $P2_1/c$-8),
  the most stable structure changes as $C2/c$-24 $\to$ $Cmca$-12 $\to$ $I4_{1}/amd$.
  This result qualitatively agree with the previous FNDMC study~\cite{2015MCM}.
  Our newly found $P2_{1}/c$-8 structure is predicted more stable than the structures above at 400 and 500~GPa.
  Therefore, we propose $P2_{1}/c$-8 as the new candidate structure of the phase H$_2$-PRE,
  which was experimentally found to exist from 360 to 495~GPa.
  The earlier candidate structure of H$_2$-PRE, $Cmca$-12~\cite{2007PIC,2019RPD_IFS},
  no longer appears to be stable structure according to our calculations. 
}

\vspace{2mm} 
\tadd{ 
  However, the difference of dynamic enthalpy between $P2_{1}/c$-8 
  and the second most stable structure is, at most, just 0.049(15)~mHartree/atom.
  Therefore, $C2/c$-24 or $Cmca$-12 can be predicted comparably stable as $P2_{1}/c$-8,
  by taking into account anharmonic contributions to ZPE~\cite{2015DRU,2014AZA,2018MON} or 
  nuclear quantum effects~\cite{2013MORb}, or using FNDMC forces~\cite{2021KN_SSb,2021JT_JTK}.
  Nevertheless, $P21/c$-8 would remain among the most stable structures. 
}

\begin{figure}
  \centering
  \includegraphics[width=\hsize]{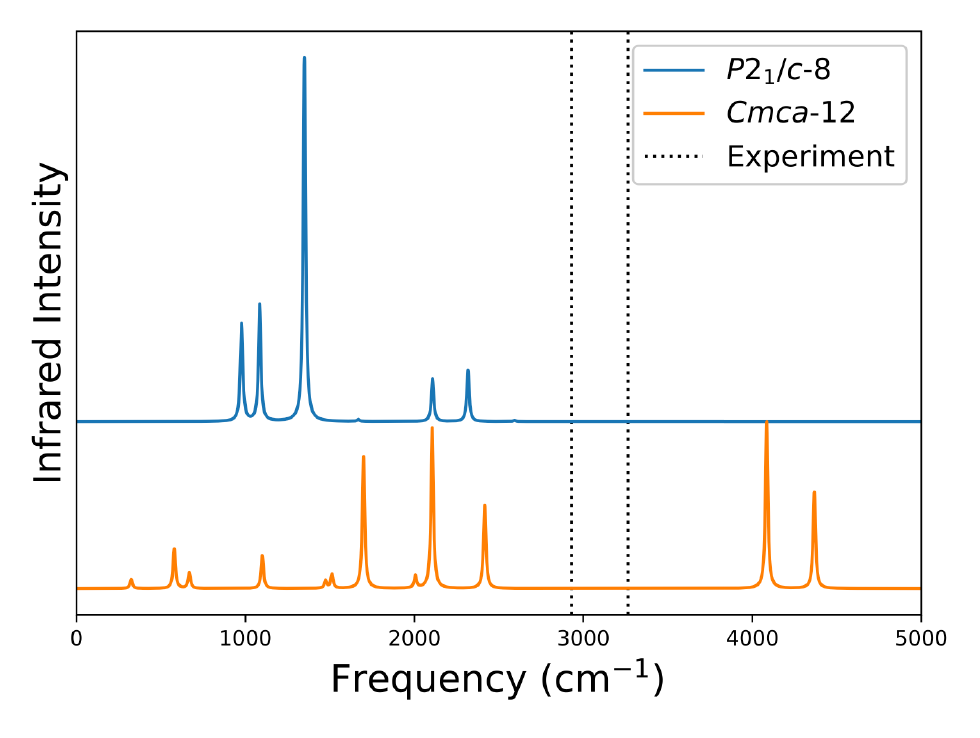}  
  \caption{
    \label{fig.ir}\ghost{fig.ir}
    The IR spectrum of $P2_{1}/c$-8 and $Cmca$-12 structures at 400~GPa.
    The vertical lines indicate the peaks found by the experiment~\cite{2019RPD_IFS}. 
  }
\end{figure}

\vspace{2mm}
\tadd{
  IR spectrum measurements revealed that the H$_2$-PRE phase has two peaks above 2800~cm$^{-1}$~\cite{2019RPD_IFS}. 
  At 400~GPa, the peaks are at 2930 and 3264 cm$^{-1}$, which are taken from Figure 2 of Ref. \onlinecite{2019RPD_IFS} 
  using WebPlotDigitizer~\cite{Rohatgi2020}. 
  Figure \ref{fig.ir} shows our predicted IR spectra obtained for $P2_{1}/c$-8 and $Cmca$-12 structures
  with vdW-DF functional compared to the peaks observed experimentally~\cite{2019RPD_IFS}. 
  $Cmca$-12 structure has two peaks around 4000~cm$^{-1}$ in consistent with the previous theoretical works
  ~\cite{2013SA_WMCF,2012CJP_RJN}.
  Existence of the two peaks is a reason why $Cmca$-12 has been considered promising as the H$_2$-PRE structure.
  Our $P2_{1}/c$-8 also has two peaks above 2000~cm$^{-1}$ similar to Cmca-12 but with a lower formation energy.
  Therefore, we propose that $P2_{1}/c$-8 is a more promising candidate structure of the H$_2$-PRE phase. 
}

\begin{figure}
  \centering
  {\Large (a) side view}  
  \includegraphics[width=\hsize]{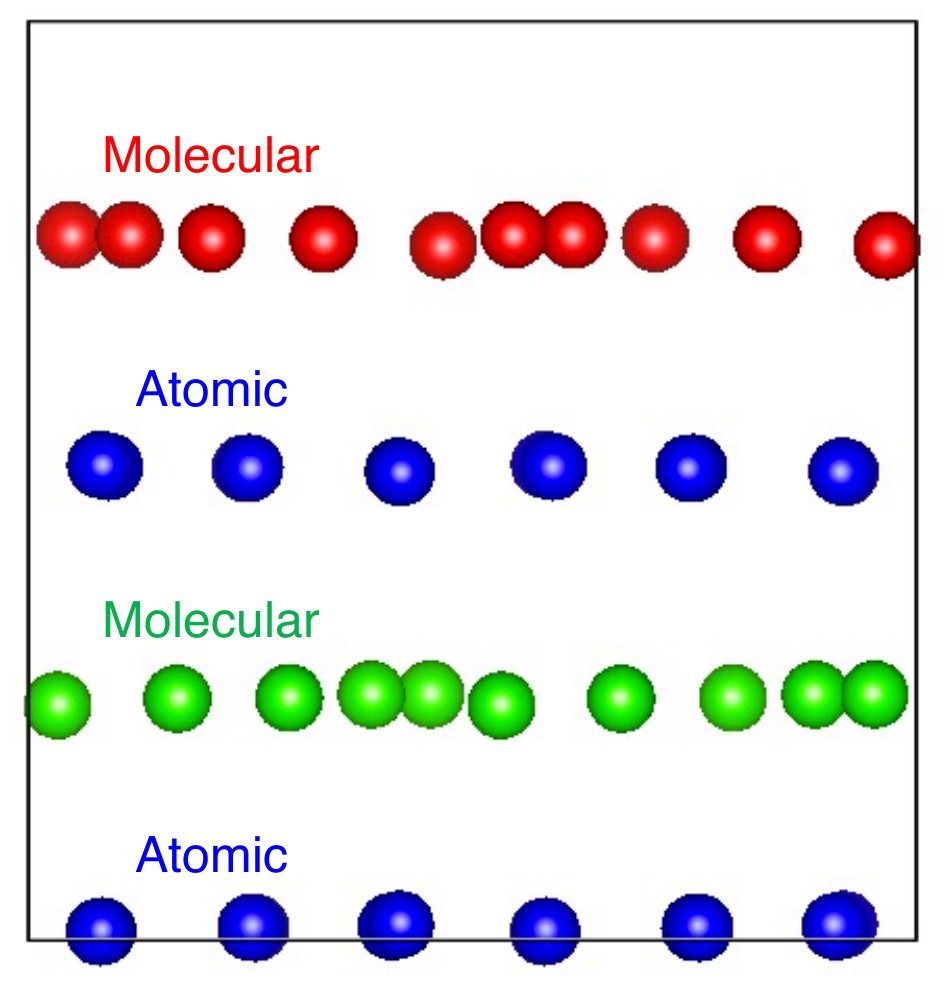}
  {\Large (b) top view}  
  \includegraphics[width=\hsize]{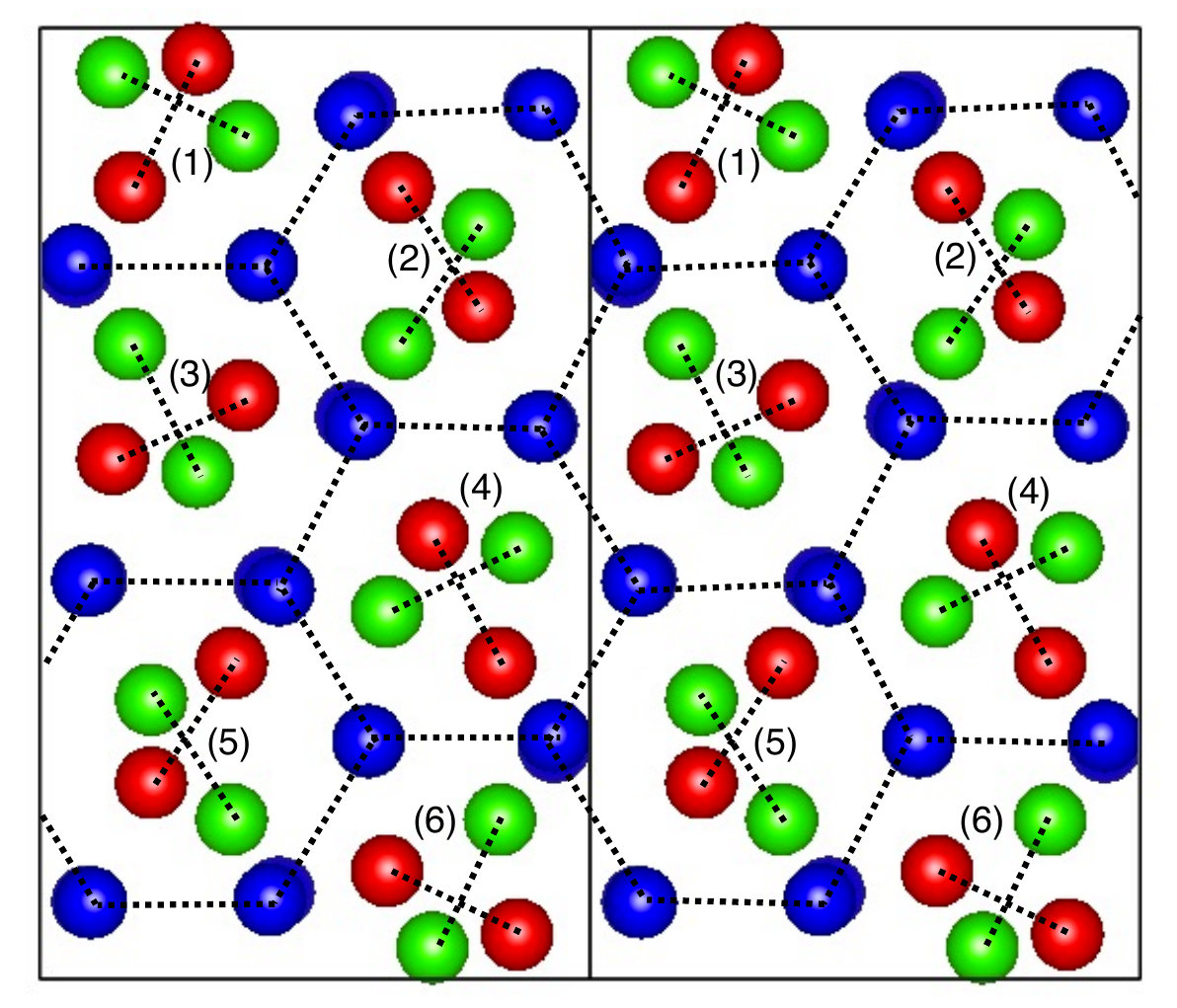}  
  \caption{
    \label{fig.pca21_structure}\ghost{fig.pca21\_structure}
    Crystal structure of $Pca2_{1}$.
    The hydrogen atoms in the atomic crystal layers are shown in blue.
    The hydrogen atoms in the molecular crystal layers are shown in red or green.
    The red and green molecular layers are different by 180$^{\circ}$ around the $c$-axis. 
  }
\end{figure}

\vspace{2mm}
\tadd{
  The structure of $P2_1/c$-8 is similar to $Pca2_{1}$ found as a candidate structure of the phase V~\cite{2018MON}.   
  The structure of $Pca2_{1}$ is shown in Figure \ref{fig.pca21_structure}.
  This structure consists of the hexagonal atomic crystal layers and the molecular layers,
  where the molecules are in the center of the hexagonal rings.
  The pronounced difference between $P2_{1}/c$-8 and $Pca2_{1}$ is the periodicity in the $c$-axis direction. 
  Only for $Pca2_{1}$, the neighboring molecular layers are rotationally different by 180$^{\circ}$ around the $c$-axis.
  In addition, $P2_1/c$-8 is more symmetric also in the $ab$-plane direction than $Pca2_{1}$.
  In a molecular layer of a periodic cell, there are two types of molecular angles for $P2_{1}/c$-8
  and eight types of molecular angles for $Pca2_{1}$.
  Over all, $P2_1/c$-8 structure is more translational symmetric than $Pca2_{1}$ structure.
  We discuss structural relationship between $P2_1/c$-8 and $C2/c$-24 or $I4_1/amd$ in the Appendix.
}

\section{Conclusion}
\label{sec.conclusion}
We performed a new structural search for solid-phase hydrogen 
at zero temperature in a high-pressure region, from 400 to 600~GPa. 
After obtaining candidate structures predicted using the PSO algorithm, 
we compared their enthalpies while considering the electron-correlation effects
using the FNDMC method with zero-point energy corrections evaluated
within the harmonic approximation of a phonon spectra.
\tadd{
  We found 10 candidate structures in our structural search.
  Among the structures,
  $Pbam$-8 is predicted to have a comparatively low static enthalpy by FNDMC
  with the previously reported structures.
  However, $Pbam$-8 has imaginary phonon modes.
  We relaxed the structure along the direction of the imaginary mode
  at the $\mathrm{\Gamma}$ point and obtained a new structure, $P2_{1}/c$-8.
  This structure has negligible  population of imaginary modes at 400 and 500~GPa,
  which we attribute to numerical error or anharmonic effects. 
  Our dynamic enthalpy evaluation given as the sum of
  the FNDMC static enthalpy and ZPE predicted by DFT
  showed that $P2_{1}/c$-8 is the most stable at 400 and 500~GPa. 
  In addition, we found that the predicted IR spectrum of $P2_{1}/c$-8
  qualitatively reproduces observed peaks for the pre-metallic phase, H$_2$-PRE~\cite{2019RPD_IFS}.
  Therefore, we propose $P2_{1}/c-8$ as the new candidate structure of the H$_2$-PRE phase,
  which was experimentally found to exist from 360 to 495~GPa~\cite{2019RPD_IFS}.
}

\section{Appendix}
\label{sec.appendix}\ghost{sec.appendix}
\begin{figure}
  \centering
  {\Large (a) side view}  
  \includegraphics[width=\hsize]{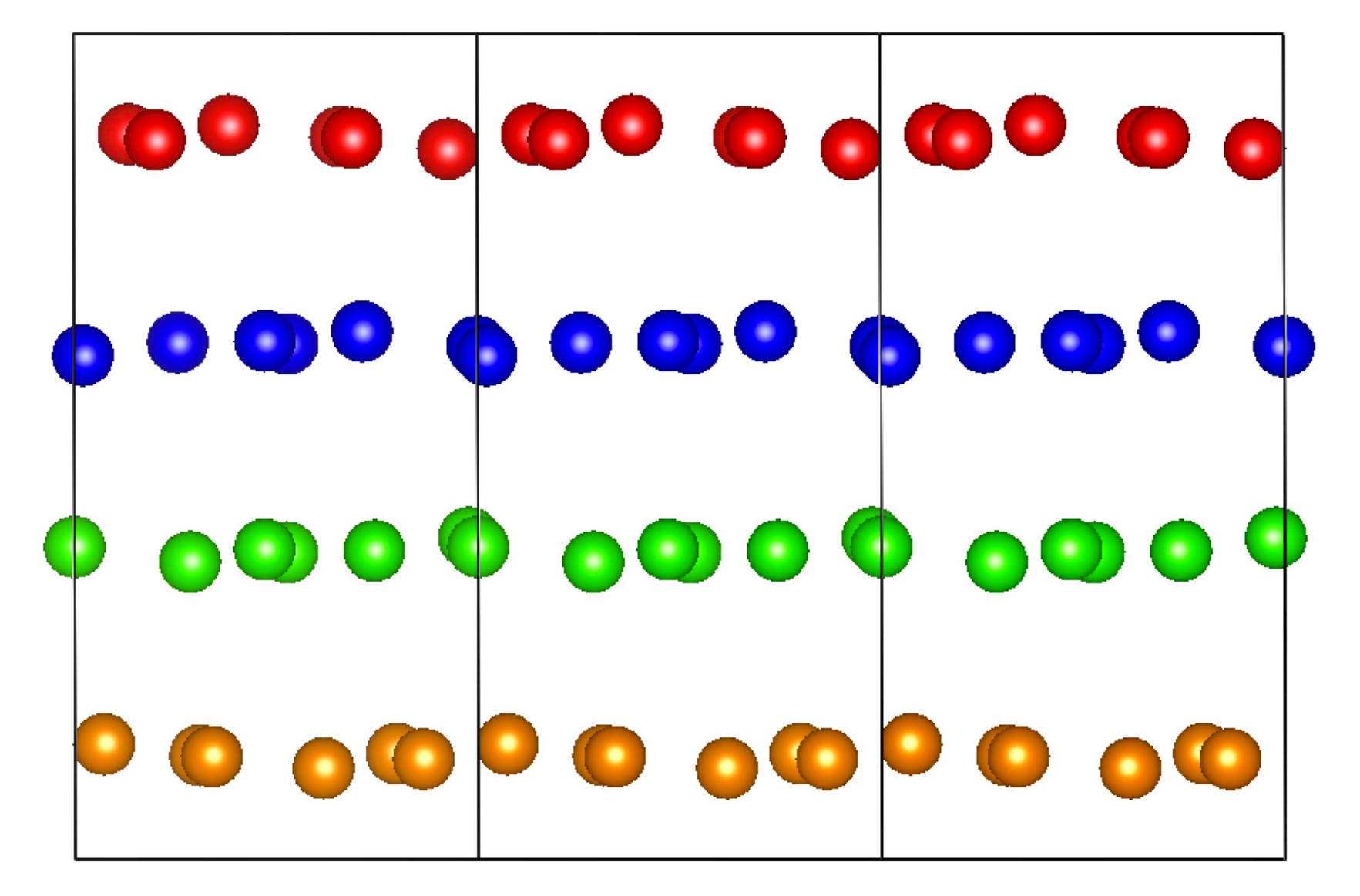}
  {\Large (b) top view}  
  \includegraphics[width=\hsize]{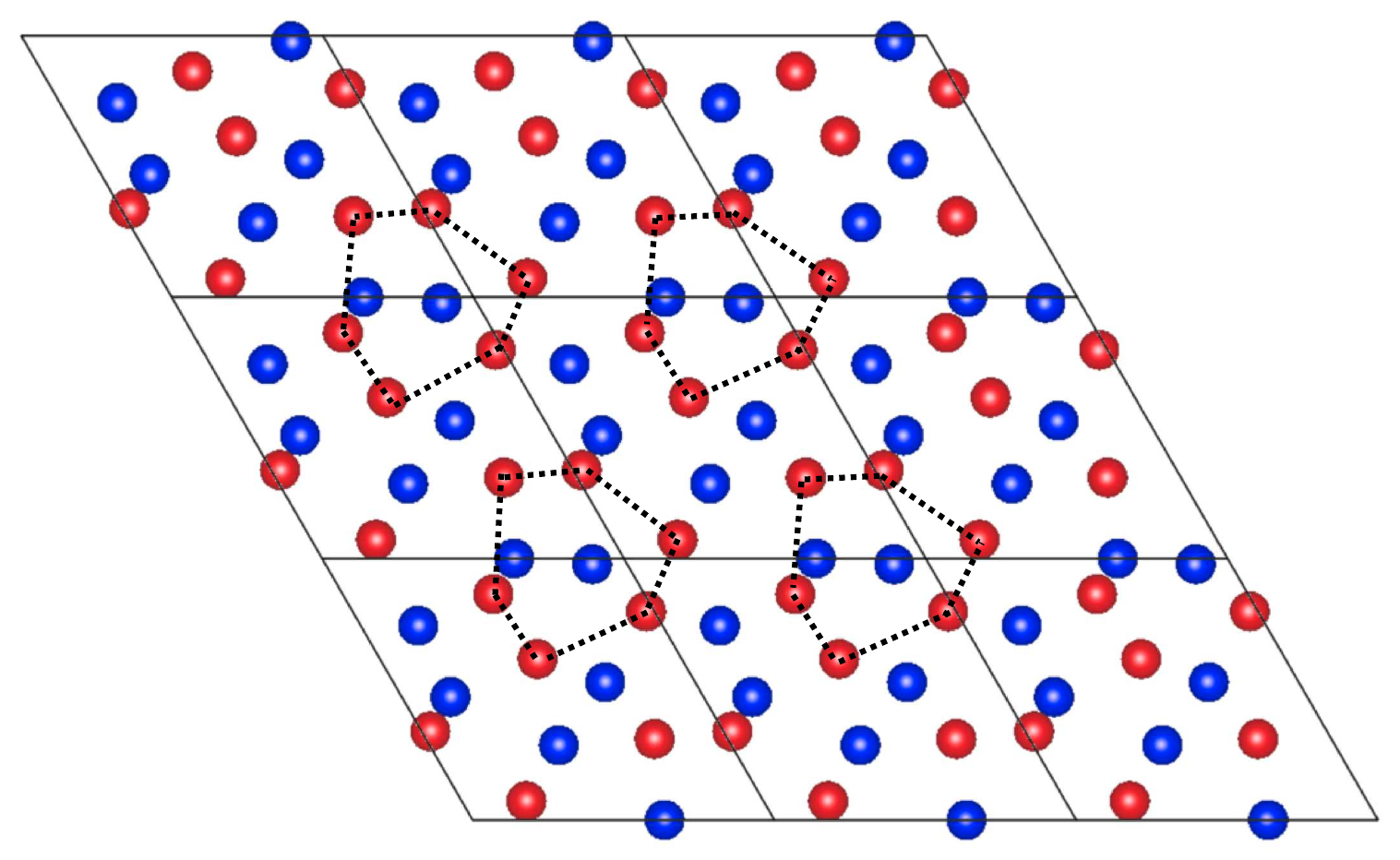}  
  \caption{
    \label{fig.c2c24_structure}\ghost{fig.c2c24\_structure}
    Crystal structure of $C2/c$-24.
    Equivalent four layers are in a unit cell.
    They have angular and translational differences.
    Only the first and second layers are shown in the lower figure.     
  }
\end{figure}
\begin{figure}
  \centering
  \includegraphics[width=\hsize]{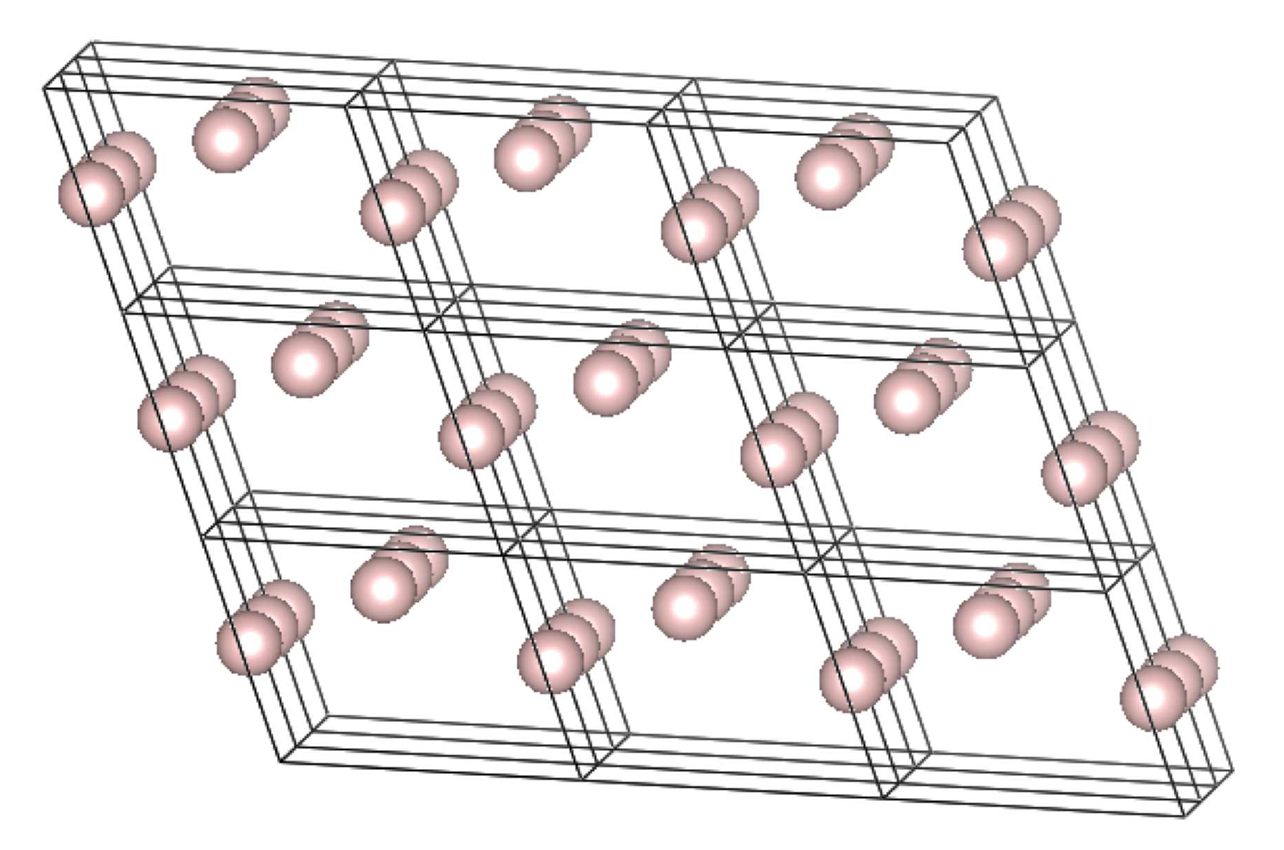}
  \caption{
    \label{fig.i41amd_structure}\ghost{fig.i41amd\_structure}
    Crystal structure of $I4_{1}/amd$. 
  }
\end{figure}

\vspace{2mm}
\tadd{
  The structure of $P2_1/c$-8 is slightly similar to that of $C2/c$-24
  found as a candidate structure of the phase III~\cite{2007PIC}.
  The structure of $C2/c$-24 is shown in Figure \ref{fig.c2c24_structure}.
  The unit cell of $C2/c$-24 consists of four equivalent molecular layers
  with angular and translational differences.
  Each layer consists of distorted hexagonal lattices. 
  Figure \ref{fig.c2c24_structure}(b) shows only two neighboring layers.
  A molecule in the blue layer is located in the center of
  a distorted hexagonal lattice, similarly to $P2_1/c$-8.
  However, the molecule in the blue layer is also a member of
  a hexagonal lattice in the blue layer.
  Compared with $C2/c$-24, $P2_1/c$-8 is regarded that
  hydrogen atoms in one layer move to form strict hexagonal lattices
  and ones in another layer move to form isolated molecules.
  On the other hand, for a candidate structure of the metallic hydrogen,
  $I4_{1}/amd$, shown in Figure \ref{fig.i41amd_structure} we cannot find any similarity
  with the other structures.
}

%

\section{Acknowledgments}
\tadd{We thank Prof. Dr. Yanming Ma, Dr. Bartomeu Monserrat, and Dr. Kosuke Nakano for the useful discussions.}
We used VESTA \cite{2011KM_FI} to draw the crystal structures. 
This work was supported by the US Department of Energy, Office of Science,
Basic Energy Sciences, Materials Sciences and Engineering Division.
We acknowledge computational resources provided by the Oak Ridge Leadership
Computing Facility at Oak Ridge National Laboratory,
which is a user facility of the Office of Science of the US Department of Energy
under Contract No. DE-AC05-00OR22725, and by the Compute and Data Environment
for Science (CADES) at Oak Ridge National Laboratory.
We also acknowledge computational resources provided 
the Research Center for Advanced Computing Infrastructure (RCACI) at JAIST.
T.I. is grateful for financial suport from
Grant-in-Aid for JSPS Research Fellow (18J12653).
\tadd{K.H. is grateful for financial support from 
  MEXT-KAKENHI (JP16H06439, JP19K05029, JP19H05169, and JP21K03400),
  and the Air Force Office of Scientific Research
  (Award Numbers: FA2386-20-1-4036).}
R.M. is grateful for financial supports from 
MEXT-KAKENHI (JP19H04692 and JP16KK0097), 
FLAGSHIP2020 (project nos. hp190169 and hp190167 at K-computer), 
the Air Force Office of Scientific Research 
(AFOSR-AOARD/FA2386-17-1-4049;FA2386-19-1-4015), 
and 
\tadd{JSPS Bilateral Joint Projects (JPJSBP120197714).}
\bibliography{references}
\end{document}